\documentclass[10pt,twocolumn,letterpaper]{article}

\usepackage[pagenumbers]{cvpr} 

\usepackage{graphicx}
\usepackage{amsmath}
\usepackage{amssymb}
\usepackage{booktabs}
\usepackage{pifont}

\usepackage[pagebackref,breaklinks,colorlinks]{hyperref}

\usepackage[capitalize]{cleveref}
\crefname{section}{Sec.}{Secs.}
\Crefname{section}{Section}{Sections}
\Crefname{table}{Table}{Tables}
\crefname{table}{Tab.}{Tabs.}

\usepackage{bbding}
\usepackage{soul}
\usepackage{bibunits}
\usepackage{bm}
\usepackage{array}
\usepackage{xcolor}
\usepackage{pdflscape}
\usepackage{makecell}
\usepackage{framed,multirow}
\usepackage{threeparttable}
\usepackage{amssymb}
\usepackage{pifont}
\usepackage{algorithm}
\usepackage{layouts}
\usepackage{listings}
\usepackage{pythonhighlight}
\makeatletter
\newcommand\footnoteref[1]{\protected@xdef\@thefnmark{\ref{#1}}\@footnotemark}
\makeatother

\newcolumntype{P}[1]{>{\centering\arraybackslash}p{#1}}
\newlength\savewidth\newcommand\shline{\noalign{\global\savewidth\arrayrulewidth
  \global\arrayrulewidth 0.8pt}\hline\noalign{\global\arrayrulewidth\savewidth}}
\newcommand{\cmark}{\ding{51}}%
\def\arrvline{\hfil\kern\arraycolsep\vline\kern-\arraycolsep\hfilneg}

\definecolor{Highlight}{HTML}{39b54a}  

\newenvironment{jlred}{\color{red}}{}

\def\cvprPaperID{1213} 
\def\confName{CVPR}
\def\confYear{2023}

\begin{document}

\title{Label-Free Liver Tumor Segmentation}

\author{Qixin~Hu\textsuperscript{1} \quad Yixiong~Chen\textsuperscript{2} \quad Junfei~Xiao\textsuperscript{3} \quad Shuwen~Sun\textsuperscript{4}\\Jieneng~Chen\textsuperscript{3} \quad Alan~Yuille\textsuperscript{3} \quad Zongwei~Zhou\textsuperscript{3,}\thanks{Corresponding author: Zongwei Zhou (\href{mailto:zzhou82@jh.edu}{zzhou82@jh.edu})}\\[2.5mm]
\textsuperscript{1}Huazhong University of Science and Technology\\
\textsuperscript{2}The Chinese University of Hong Kong -- Shenzhen \\
\textsuperscript{3}Johns Hopkins University\\
\textsuperscript{4}The First Affiliated Hospital of Nanjing Medical University\\ [1.5mm]
{\small Code and Visual Turing Test:~\href{https://github.com/MrGiovanni/SyntheticTumors}{https://github.com/MrGiovanni/SyntheticTumors}}
}

\maketitle

\begin{abstract}

We demonstrate that AI models can accurately segment liver tumors without the need for manual annotation by using synthetic tumors in CT scans. Our synthetic tumors have two intriguing advantages: (I) realistic in shape and texture, which even medical professionals can confuse with real tumors; (II) effective for training AI models, which can perform liver tumor segmentation similarly to the model trained on real tumors---this result is exciting because no existing work, using synthetic tumors only, has thus far reached a similar or even close performance to real tumors. This result also implies that manual efforts for annotating tumors voxel by voxel (which took years to create) can be significantly reduced in the future. Moreover, our synthetic tumors can automatically generate many examples of small (or even tiny) synthetic tumors and have the potential to improve the success rate of detecting small liver tumors, which is critical for detecting the early stages of cancer. In addition to enriching the training data, our synthesizing strategy also enables us to rigorously assess the AI robustness.

\end{abstract}

\section{Introduction}
\label{sec:introduction}

\begin{figure*}[t]
\centerline{\includegraphics[width=1.0\textwidth]{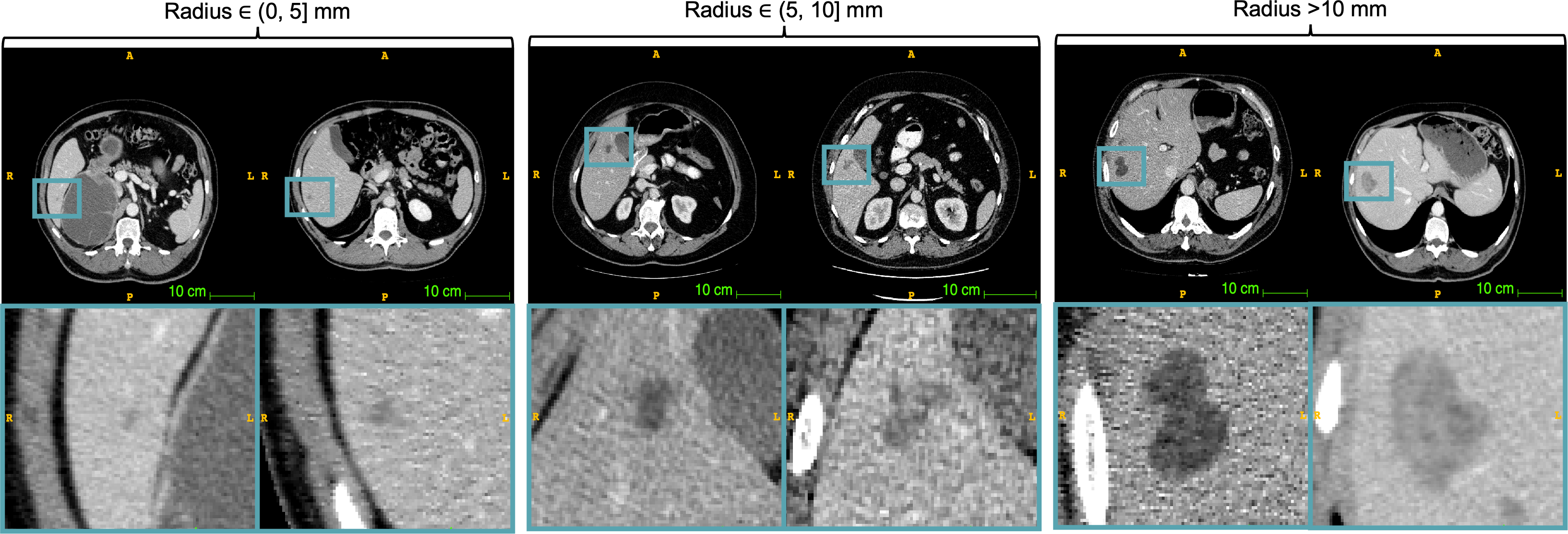}}
    \caption{
    [Better viewed in color and zoomed in for details] \textit{Can you tell which liver tumors are real and which are fake?} The answers are provided in Appendix. We have recruited two medical professionals with at least six years of experience to distinguish fake tumors, generated by our method, from the real ones (namely, Visual Turing Test). Our synthetic tumors have passed the Visual Turing Test on both two medical professionals ($<$50\% fake tumors were picked out). More importantly, using our label-free synthetic tumors, AI models can segment real tumors with performance similar to the AI models trained on real tumors with expensive, detailed, per-voxel annotation.
    }
\label{fig:teaser}
\end{figure*}

Artificial intelligence (AI) has dominated medical image segmentation~\cite{zhou2019unet++,isensee2021nnu,hatamizadeh2022unetr,zhou2019models,zhou2021models}, but training an AI model (\eg, U-Net~\cite{ronneberger2015u}) often requires a large number of annotations. Annotating medical images is not only expensive and time-consuming, but also requires extensive medical expertise, and sometimes needs the assistance of radiology reports and biopsy results to achieve annotation accuracy~\cite{zhou2021review,erickson2021imaging,rubin2021biomedical,zhou2021towards,wang2021development,zhou2022interpreting}. Due to its high annotation cost, only roughly 200 CT scans with annotated liver tumors are publicly available (provided by LiTS~\cite{bilic2019liver}) for training and testing models.

\begin{figure}[t]
\centerline{\includegraphics[width=1.0\columnwidth]{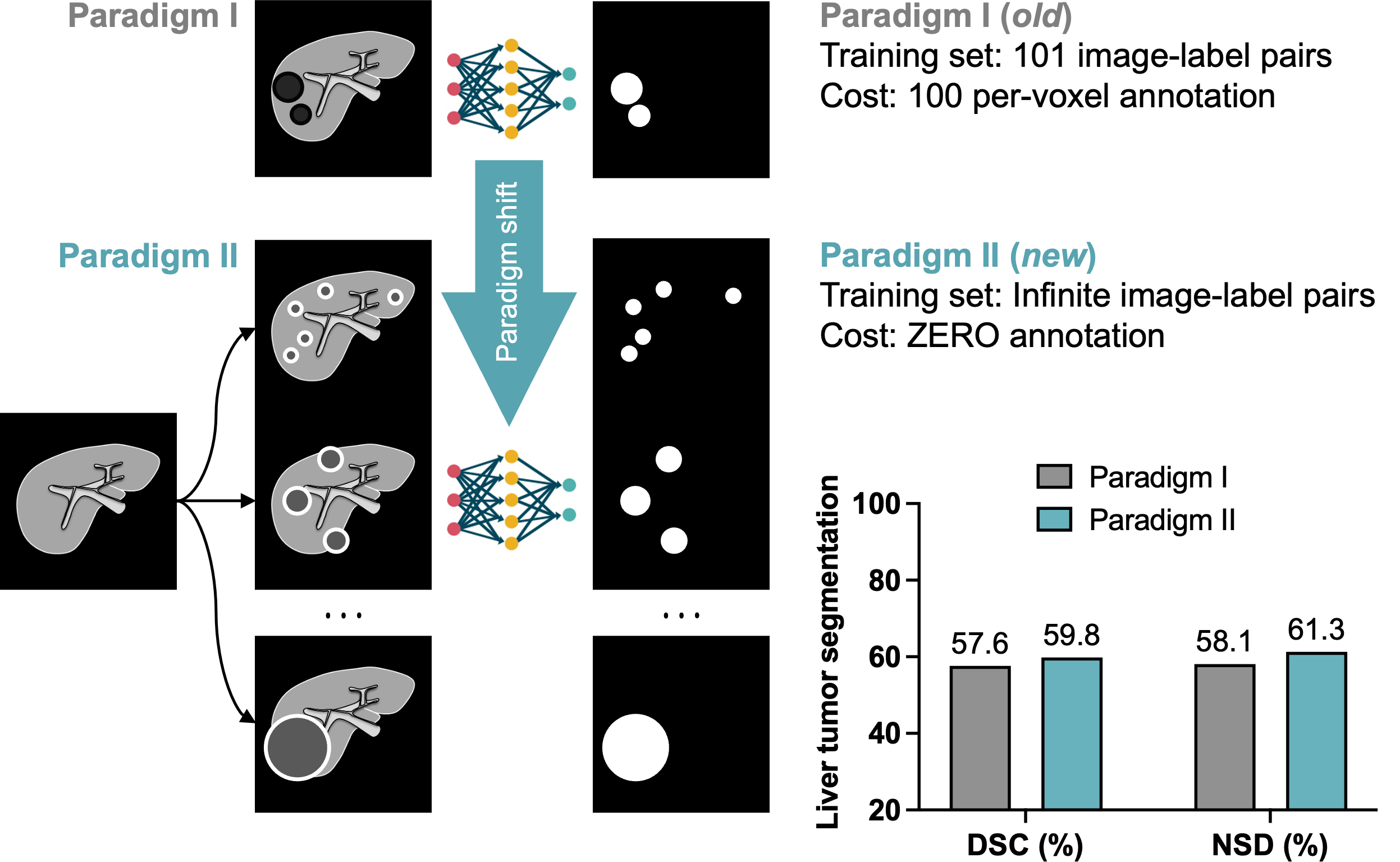}}
    \caption{
    The paradigm shift from \textit{label-intensive} to \textit{label-free} tumor segmentation in this work. AI trained on synthetic tumors can segment liver tumors as accurately as AI trained on real tumors. The performance is measured on real tumors using Dice Similarity Coefficient (DSC) and Normalized Surface Distance (NSD).
    }
\label{fig:paradigm_shift}
\end{figure}

To minimize annotation expenses, generating synthetic tumors is an emerging research topic. Early attempts include, but only limited to, synthesizing COVID-19 infections~\cite{yao2021label,lyu2022pseudo}, lung nodules~\cite{han2019synthesizing} abdominal tumors~\cite{jin2021free}, diabetic lesions~\cite{wang2022anomaly}, and brain tumors~\cite{wyatt2022anoddpm}. 
However, the synthetic tumors in those studies appear very different from the real tumors; due to this, AI models trained using synthetic tumors perform significantly worse than those trained using real tumors. 
\textit{What makes synthesizing tumors so hard?} 
There are several important factors: shape, intensity, size, location, and texture.
In this paper, we handcraft a strategy to synthesize liver tumors in abdominal CT scans.
Our key novelties include \textit{(i)} location without collision with vessels, \textit{(ii)} texture with scaled-up Gaussian noise, and \textit{(iii)} shape generated from distorted ellipsoids.
These three aspects are proposed according to the clinical knowledge of liver tumors (detailed in~\S\ref{sec:clinical_knowledge}).
The resulting synthetic tumors are realistic---even medical professionals usually confuse them with real tumors in the visual examination (\figureautorefname~\ref{fig:teaser}; \tableautorefname~\ref{tab:turing_test_result}). In addition, the model trained on our synthetic tumors achieves a Dice Similarity Coefficient (DSC) of 59.81\% for segmenting real liver tumors, whereas AI trained on real tumors obtains a DSC of 57.63\%  (\figureautorefname~\ref{fig:paradigm_shift}), showing that synthetic tumors have the potential to be used as an alternative to real tumors in training AI models.

These results are exciting because using synthetic tumors \textit{only}, no previous work has thus far reached a similar (or even close) performance to the model trained on real tumors~\cite{hu2022synthetic}. Moreover, our synthesizing strategy can exhaustively generate tumors with desired locations, sizes, shapes, textures, and intensities, which are not limited to a fixed finite-size training set (the well-known limitation of the conventional training paradigm~\cite{yuille2021deep}). 
For example, it is hard to collect sufficient training examples with small tumors. It is because early-stage tumors may not cause symptoms, which can delay detection, and these tumors are relatively small and exhibit subtle abnormal textures that make it difficult for radiologists to manually delineate the tumor boundaries. In contrast, our synthesis strategy can generate a large number of examples featuring small tumors. 
The key \textbf{contribution} of ours is a synthetic tumor generator, which offers five advantages as summarized below.

\begin{figure*}[t]
\centerline{\includegraphics[width=1.0\linewidth]{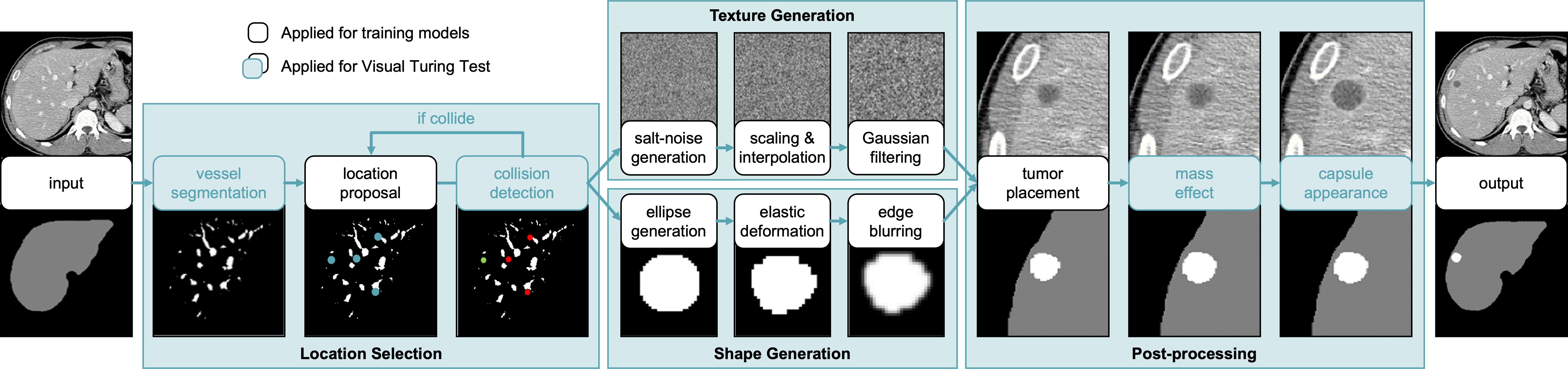}}
\caption{
\textbf{Liver tumor generation.} After randomly selecting a location that avoids the vessels, we generate a Gaussian texture and deformed ellipsoidal shape for a tumor. Then, the texture and shape are combined and placed in the selected location. In addition, we take another two post-processing steps to make a generated tumor more realistic: (1) tumor edge expansion by local scaling warping; (2) capsular generation by brightening the tumor edge. Four steps, in light green, are only used for Visual Turing Test (not for training).
}
\label{fig:method}
\end{figure*}

\begin{enumerate}
    \item The synthesis strategy embeds medical knowledge into an executable program, enabling the generation of realistic tumors through the collaboration of radiologists and computer scientists (\S\ref{sec:visual_turing_test}; \tableautorefname~\ref{tab:turing_test_result}; \figureautorefname~\ref{fig:method}).
    
    \item The entire training stage requires no annotation cost, and the resulting model significantly outperforms previous unsupervised anomaly segmentation approaches and tumor synthesis strategies (\S\ref{sec:outperform_baseline}; \tableautorefname~\ref{tab:synt_real_result}).

    \item The AI model trained on synthetic tumors can achieve similar performance to AI models trained on real tumors with per-voxel annotation in real tumors segmentation, and can be generalized to CT scans with healthy liver and scans from other hospitals {\jlred }(\S\ref{sec:generalization_sim2real}; \figureautorefname~\ref{fig:tiny_small_base}).
    
    \item The synthesis strategy can generate a variety of tumors for model training, including those at small, medium, and large scales, and therefore have the potential to detect small tumors and facilitate the early detection of liver cancer (\S\ref{sec:small_tumor_detection}; \figureautorefname~\ref{fig:performance_tumor_size}).
    
    \item The synthesis strategy allows for straightforward manipulation of parameters such as tumor location, size, texture, shape, and intensity, providing a comprehensive test-bed for evaluating AI models under out-of-distribution scenarios (\S\ref{sec:robustness_benchmark}; \figureautorefname~\ref{fig:ood_evaluation}).
\end{enumerate}

These results have the potential to stimulate a shift in the tumor segmentation training paradigm, as illustrated in~\figureautorefname~\ref{fig:paradigm_shift}, from \textit{label-intensive} to \textit{label-free} AI development for tumor segmentation. 
Our ultimate goal is to train AI models for tumor segmentation without using manual annotation---this study makes a significant step towards it.

\section{Related Work}
\label{sec:related_work}

\noindent\textbf{\textit{Unsupervised anomaly segmentation.}}
Anomaly segmentation is a challenging application area, especially for medical diagnosis~\cite{kermany2018identifying,xiang2021painting} and industrial defect detection~\cite{bergmann2019mvtec,huang2022self}. Compared with their supervised counterparts, unsupervised methods raises more attention for their low cost and scalability. The general unsupervised anomaly detection setting is to train with normal samples only, without any anomalous data, and no image-level annotation or pixel-level annotation is provided\cite{roth2022towards,schlegl2019f,kingma2013auto}. Under the unsupervised setting, some previous works use self-organizing maps for unsupervised anomaly detection~\cite{munoz1998self,li2021anomaly} and Huang~\etal~\cite{huang2022self} introduced gradient magnitude similarity and structured similarity index losses in addition to mean square error to compute the loss of image reconstruction. Evidenced in~\tableautorefname~\ref{tab:synt_real_result}, our label-free synthetic tumors achieve a significantly better performance in unsupervised tumor segmentation than some of the most recent work in this area. 

\smallskip\noindent\textbf{\textit{Tumor synthesis.}}
Successful works about tumor synthesis include polyp detection from colonoscopy videos~\cite{shin2018abnormal}, COVID-19 detection from Chest CT~\cite{yao2021label,lyu2022pseudo}, diabetic lesion detection from retinal images~\cite{wang2022anomaly}, cancer detection from fluorescence microscopy images~\cite{horvath2022metgan}, and brain tumor detection from MRI~\cite{wyatt2022anoddpm}. 
However, these works are restricted to the types of tumors, and other diseases, that are fairly easy to visually identify in CT scans. Most recently, Zhang~\etal~\cite{zhang2023self} synthesized liver and brain tumors for pre-training and adapted the model to tumor segmentation within the same organ under a low-annotation regime. The manually crafted ``counterfeit'' tumors in the related work appear very differently from real tumors. As a result, AI algorithms, trained on synthetic tumors, may work well in detecting synthetic tumors in the test set but fail to recognize the actual tumors (evidenced in~\tableautorefname~\ref{tab:synt_real_result}).
We tackle these limitations by integrating radiologists in the tumor synthesis study for feedback (\S\ref{sec:clinical_knowledge}). This enables us to understand deeply about tumors, and in turn, benefit in developing AI algorithms to segment them more accurately. 

\smallskip\noindent\textbf{\textit{Generalization from synthetic to real domains.}}
The problem of domain generalization was introduced~\cite{blanchard2011generalizing} for zero-shot adaptation to data with a domain gap. Specifically, the goal is to make a model using data from a single or multiple related source domain(s) while achieving great generalization ability well to any target domain(s)~\cite{zhou2022domain, wang2022generalizing}. In this paper, evaluating the model generalization ability on real data is of great importance to justify whether our tumor generator is powerful enough. Domain generalization has been widely studied in multiple computer vision tasks like object recognition~\cite{li2017deeper,feature_critic}, semantic segmentation~\cite{volpi2019addressing,yue2019domain} and medical imaging~\cite{liu2020ms,liu2020shape}. As deep learning models are data hungry and annotated data are very expensive, how to train a model with synthetic data but generalize well to real data has been targeted in some previous works~\cite{fang2013unbiased, chen2020automated, chen2021contrastive} and some datasets~\cite{peng2017visda,ganin2015unsupervised,ros2016synthia,cordts2016cityscapes, richter2016playing} are created for benchmark and further exploring.
While previous works focus on preserving the transferable knowledge learned from synthetic data, our paper aims to prove that our tumor generator is powerful enough to generate tumors with reasonable domain gap and that our model has outstanding generalization ability to detect real tumors (detailed in~\Cref{sec:generalization_sim2real}).

\section{Method}\label{sec:method}


\subsection{Tumor Generation}
\label{sec:tumor_generation}

To localize the liver, we first apply the pre-trained nnU-Net\footnote{The off-the-shield, pre-trained nnU-Net~\cite{isensee2021nnu} for liver segmentation can be downloaded \href{https://zenodo.org/record/4003545/files/Task003_Liver.zip?download=1}{here}, which can achieve an average DSC of 95\% on unseen CT scans (sufficient for a coarse localization of the liver).} to the CT scans.
With a coarse location of the liver available, we then develop a sequence of morphological image-processing operations to synthesize realistic tumors within the liver (see \figureautorefname~\ref{fig:method}). 
The tumor generation consists of four steps: (1) location selection, (2) texture generation, (3) shape generation, and (4) post-processing.



\smallskip\noindent\textbf{\textit{Location selection.}} 
The first step is to select a proper location for the tumor. This step is crucial because liver tumors usually do not allow any vessels (\eg, hepatic vein, portal vein, and inferior vena cava) to pass through them. 
To avoid the blood vessels, we first conduct vessel segmentation through the voxel value thresholding~\cite{gonzalez2009digital}. The segmented vessel mask is given by the following equation:
\begin{equation}
v(x,y,z)=
\left\{
 \begin{array}{lr}
 1,~f'(x,y,z)>T, ~l(x,y,z)=1 \\
 0,~\text{otherwise}~~~~~~~~~~~~~~~~~~~~~~~~~~~~~~~
 \end{array},
\right.
\end{equation}
where $f'(x,y,z)$ is the smoothed CT scan, $f'(x,y,z)=f(x,y,z)\otimes g(x,y,z;\sigma_a)$,
by applying a Gaussian filter $g(x,y,z;\sigma_a)$ with standard deviation $\sigma_a$ to the original CT scans $f(x,y,z)$; $\otimes$ is the standard image filtering operator. Smoothing can effectively eliminate noise caused by CT reconstruction. The threshold $T$ is set to a value slightly greater than the mean Hounsfield Unit (HU) of the liver.
\begin{equation}
    T = \overline{f(x,y,z)\odot l(x,y,z)} + b,
\end{equation}
where $l(x,y,z)$ is the liver mask (background=0, liver=1), $\odot$ is point-wise multiplication, and $b$ is a hyperparameter.

With the vessel mask, one can detect whether a chosen location is at risk of making a tumor collide with vessels. After proposing a random location $(X,Y,Z)\in \{x,y,z~|~ l(x,y,z)=1\}$, we conduct the collision detection by judging whether there are blood vessels within the range of tumor radius $r$. If $\exists ~v(x,y,z)=1, \forall ~x\in[X-r,X+r],~y\in[Y-r,Y+r],~z\in[Z-r, Z+r]$, there is a risk of collision, so the location needs to be re-selected. This process iterates until a tumor location $(x_t,y_t,z_t)$ without collision is found. With the desirable tumor location, we are able to generate the tumor texture and shape.

\smallskip\noindent\textbf{\textit{Texture generation.}} The HU values of liver and tumor textures follow the Gaussian distributions. To obtain realistic tumor textures, we first generate a 3D Gaussian noise with the predefined mean HU intensity $\mu_t$ and the same standard deviation $\sigma_p$ as the hepatic parenchyma (the liver area excluding vessels), $T(x,y,z)\sim \mathcal{N}(\mu_t, \sigma_p)$.
Since the random Gaussian noise is usually too sharp as the texture for the tumor, we soften the texture by scaling it up with spline interpolation of the order 3 (cubic interpolation) on $x,y,z$ directions. The scaled-up texture is denoted as $T'(x,y,z)$ in this work, we want it exhibits graininess close to the hepatic parenchyma. The scaling factor $\eta\in [1,\infty)$ determines how rough the generated grain feels. $\eta=1$ means the Gaussian texture is not scaled, resulting in large value fluctuation between adjacent voxels. Larger $\eta$ brings greater graininess, which may be similar to the real tumor texture. 
Finally, considering the tomography imaging quality, we further blur the texture with Gaussian filter $g(x,y,z;\sigma_b)$
\begin{equation}
T''(x,y,z)=T'(x,y,z)\otimes g(x,y,z;\sigma_b),
\end{equation}
where $\sigma_b$ is the standard deviation. After blurring, the texture resembles those generated by real imaging.



\smallskip\noindent\textbf{\textit{Shape generation.}} Most tumors grow from the centers and gradually swell, making small tumors (\ie, $r<20mm$) nearly spherical. This motivates us to generate tumor-like shapes with ellipsoids. We randomly sample the half-axis lengths of the ellipsoid for $x,y,z$ directions from a uniform distribution $U(0.75r, 1.25r)$, and place the generated ellipsoid mask centered at $(x_t,y_t,z_t)$. 
For a generated ellipsoid tumor mask $t(x,y,z)$ (background=0, tumor=1), and with the same shape as the scanning volume $f(x,y,z)$), elastic deformations~\cite{ogden1997non,ronneberger2015u}, controlled by $\sigma_e$, are applied to enrich its diversity. The deformed tumor mask is more similar to the naturally grown tumors in appearance than simple ellipsoids. In addition, it can also improve the model's robustness by learning the shape-semantic invariance. The deformed tumor mask is denoted as $t'(x,y,z)$. 
In order to make the transition between the generated tumor and the surrounding liver parenchyma more natural, we finally blur the mask by applying a Gaussian filter $g(x,y,z;\sigma_c)$ with the standard deviation $\sigma_c$. To be more specifically, we obtain blur shape $t''(x,y,z)=t'(x,y,z)\otimes g(x,y,z;\sigma_c)$.

\smallskip\noindent\textbf{\textit{Post-processing.}} The first step of post-processing is placing the tumor on the scanning volume $f(x,y,z)$ and corresponding liver mask $l(x,y,z)$. Assuming the tumor mask array $t''(x,y,z)$ and the texture array $T''(x,y,z)$ have the same shape as $f(x,y,z)$ and $l(x,y,z)$. We can obtain new scanning volume with tumor through equation
\begin{align}
    f'(x,y,z) =& (1-t''(x,y,z))\odot f(x,y,z) + \\
    & t''(x,y,z)\odot T''(x,y,z). \notag
\end{align}
For the new mask with the tumor (bg=0,liver=1,tumor=2), it can be synthesized with $l'(x,y,z) = l(x,y,z) + t''(x,y,z)$.
After placing the tumor, we adopt another two steps to make the generated tumor more realistic to medical professionals. They aim to simulate mass effect and the capsule appearance, respectively. 
Mass effect means the expanding tumor pushes its surrounding tissue apart. If the tumor grows large enough, it will compress the surrounding blood vessels to make them bend, or even cause the edge of the nearby liver to bulge. Local scaling warping~\cite{gustafsson1993interactive} is chosen in this work to implement mass effect. It remaps pixels in a circle to be nearer to the circumference. For a pixel with a distance $\gamma$ to the circle center, the remapped pixel distance $\gamma'$ is
\begin{equation}
    \gamma' = \left (1 - \left (1 - \frac{\gamma}{\gamma_{max}}\right )^2 \cdot \frac{I}{100} \right ) \cdot \gamma,
\end{equation}
where $\gamma_{max}$ is the radius of the expanded circular area, $I\in [0,100]$ is a hyper-parameter for controlling the expansion intensity. Larger $I$ leads to stronger warping. Note that when $I=0$, the remapping reduces to the identity function $\gamma'=\gamma$. The remapping procedure is conducted on both scanning and mask volumes. After warping, they are named $f''(x,y,z)$ and $l''(x,y,z)$. The latter one (liver/tumor segmentation label) is now ready for the subsequent training.
Finally, we simulate the capsule appearance by brightening the tumor edge. The edge area can be obtained by
\begin{equation}
e(x,y,z)=
\left\{
 \begin{array}{lr}
 1,~t''(x,y,z)\in [lb,ub] \\
 0,~\text{otherwise}
 \end{array},
\right.
\end{equation}
where $lb$ and $ub$ are the lower bound and upper bound for filtering the edge from tumor mask. Then we increase HU intensity of the blurred edge area to simulate the capsule
\begin{equation}
    e'(x,y,z) = e(x,y,z) \otimes g(x,y,z;\sigma_d),
\end{equation}
\begin{equation}
    f'''(x,y,z) = f''(x,y,z) + d \cdot e(x,y,z),
\end{equation}
where $d$ is the pre-defined HU intensity difference between a tumor and its capsule. The new scanning volume $f'''(x,y,z)$ is now ready for training or Turing test. The parameters we use are shown in \tableautorefname~\ref{tab:hyper-parameter}. Visualization examples can be found in Appendix Figures~\ref{fig:extensive_samples}--\ref{fig:texture_parameters}.

\begin{table}[t]
\centering
\footnotesize

\begin{tabular}{P{0.15\linewidth}P{0.25\linewidth}|P{0.15\linewidth}P{0.25\linewidth}}
	\hline
	parameter & value & parameter & value \\ 
	\shline
	$\sigma_a$ & $0.5+0.025\sigma_p$ & $\mu_t$ & $U(30, \mu_p-10)$\\
	$\sigma_b$ & $0.6$ & $\eta$ & $U(1.1, 1.5)$ \\
	$\sigma_c$ & $U(0.6,1.2)$ & $\gamma_{max}$ & $1.3r$\\
	$\sigma_d$ & $0.8$ & $I$  & $30$ \\
	$b$ & 15 & $(lb, ub)$ & $(0.4, 0.7)$ \\
	$d$ & $120$ \\
	\hline
\end{tabular}
\caption{\textbf{Hyper-parameters.} $\mu_p, \sigma_p$ are the mean and standard deviation of the hepatic parenchyma. The values are adjusted by (1) feedback from clinicians based on the clinical prior knowledge about liver tumors (\S\ref{sec:clinical_knowledge}) and (2) visual assessment between the real and synthetic tumors (\S\ref{sec:visual_turing_test}).}
\label{tab:hyper-parameter}
\end{table}
\subsection{Clinical Knowledge about Liver Tumors}
\label{sec:clinical_knowledge}

This work focus on generating hepatocellular carcinomas (tumor grown from liver cells). After the contrast injection, the clinical examination process of liver is divided into three phases, arterial phase (30s after injection), portal venous phase (60--70s after injection), and delay phase (3min after injection). Typically, only the first two phases are used for detecting hepatocellular carcinomas, and the tumor HU intensity value in different stages distributes differently. The mean attenuation measurement of the lesions in the hepatic arterial phase was 111 HU (range, 32–207 HU), and it decreased in the portal venous phase to a mean of 106 HU (range, 36--162 HU). There was a mean difference of 26 HU (range, –44 to 146 HU) between the lesion and liver in the arterial phase. On average, the hepatocellular carcinomas measured 11 HU (range, –98 to 61 HU) less than the adjacent liver parenchyma in the portal venous phase~\cite{lee2004triple}. The distributional characteristics help us determine the generated mean tumor HU values.

The location, shape, and number of tumors depend on how severe the hepatocellular carcinomas are according to the standardized guidance of the Liver Imaging Reporting and Data System (LI-RADS)~\cite{m2021use}. Milder carcinomas usually lead to smaller, fewer spherical lesions. Only one small tumor emerges in most cases. While multi-focal lesions, which means scattered small tumors, only appear in seldom cases. Severe carcinomas usually present a satellite lesion, a large lesion surrounded by a cluster of small lesions. The large central lesion also takes on a more irregular shape than small lesions. And also, larger tumors usually display evident mass effects, accompanied by capsule appearances that separate the tumor from the liver parenchyma.




\section{Experiments}

\noindent\textbf{\textit{Datasets.}}
Detailed per-voxel annotations for liver tumors are provided in LiTS~\cite{bilic2019liver}.
The volume of liver tumors ranges from 38mm$^3$ to 349 cm$^3$, and the radius of tumors is in the range of [2, 44]mm.
We perform 5-fold cross-validation, following the same split as in Tang~\etal~\cite{tang2022self}.
An AI model (\eg U-Net) is trained on 101 CT scans with annotated liver and liver tumors.
For comparison, a dataset of 116 CT scans with healthy livers is assembled from CHAOS~\cite{kavur2021chaos} (20 CT scans), BTCV~\cite{landman2015} (47 CT scans), Pancreas-CT~\cite{TCIA_data} (38 CT scans) and health subjects in LiTS (11 CT scans). We then generate tumors in these scans on the fly, resulting in enormous image-label pairs of synthetic tumors for training the AI model. We generate five levels of tumor sizes for model training; the parameters and examples can be found in Appendix Table~\ref{tab:tumor_size_for_training} and Figure~\ref{fig:model_training_type}.


\begin{table}[t]
\centering
\footnotesize

\begin{threeparttable}
\begin{tabular}{p{0.03\linewidth}p{0.17\linewidth}|P{0.12\linewidth}P{0.12\linewidth}|P{0.12\linewidth}P{0.12\linewidth}}
    \hline
     & & \multicolumn{2}{c|}{junior professional} & \multicolumn{2}{c}{senior professional} \\
    \cline{3-6}
     & & real ($P$) & synt ($N$) & real ($P$) & synt ($N$) \\
    \shline
    \parbox[t]{0mm}{\multirow{2}{*}{\rotatebox[origin=c]{90}{truth}}} & real ($P$) & 5 & 15 & 10 & 2 \\
     & synt ($N$) & 21 & 8 & 7 & 12 \\
    \hline
\end{tabular}
\begin{tablenotes}
    \scriptsize
    \item $^1$The junior professional achieves an Accuracy, Sensitivity, and Specificity of 26.5\%, 27.6\%, and 25.0\%. One CT scan is marked \textit{unsure}.
    \item $^2$The senior professional achieves an Accuracy, Sensitivity, and Specificity of 71.0\%, 63.2\%, and 83.3\%. 19 CT scans are marked \textit{unsure}.
\end{tablenotes}
\end{threeparttable}
\caption{
    \textbf{Results of Visual Turing Test.} The test has been performed on two medical professionals  with 6-year and 15-year experience. Each professional is given 50 CT scans, some of which contain real tumors and the others contain synthetic ones. The professional can mark each CT scan as \textit{real}, \textit{synthetic}, or \textit{unsure}. ``Synt'' denotes synthetic tumors, $P$ and $N$ indicate positive and negative classes for computing Sensitivity and Specificity.
}
\label{tab:turing_test_result}
\end{table}

\begin{table*}[!ht]
    \centering
    \footnotesize

    \begin{tabular}{p{0.08\linewidth}p{0.12\linewidth}p{0.16\linewidth}P{0.16\linewidth}P{0.16\linewidth}P{0.16\linewidth}}
        \hline
        tumors & method & architecture & labeled / unlabeled CTs & DSC (\%) [95\% CI] & NSD (\%) [95\% CI] \\
        \shline
        none & PatchCore~\cite{roth2022towards} & Wide-Resnet50-2~\cite{zagoruyko2016wide} & 0 / 116 & 15.97 [11.86--20.09] & 16.43 [10.42--22.44] 
        \\
        none & f-AnoGAN~\cite{schlegl2019f}     & Customized~\cite{baur2021autoencoders} & 0 / 116 & 19.00 [13.88--24.11] & 16.94 [11.97--21.91] 
        \\
        none & VAE~\cite{kingma2013auto}        & Customized~\cite{baur2021autoencoders} & 0 / 116 & 24.63 [19.83--29.44] & 23.63 [18.44--28.83] 
        \\
        synt & Yao~\etal~\cite{yao2021label} & U-Net~\cite{ronneberger2015u} & 0 / 116 & 32.79 [28.66--36.92] & 31.28 [26.87--35.70] 
        \\
        \hline
        real & fully-supervised & U-Net & 101 / 0  & 57.51 [52.24--62.79] & 58.04 [52.56--63.52]  
        \\ 
        synt & label-free (ours) & U-Net & 0 / 116 & \textbf{59.77 [54.54--64.99]} &  \textbf{61.29 [56.12--66.47]}
        \\
        \hline
    \end{tabular}
        \caption{
        \textbf{Comparison with state-of-the-art methods, 5-fold cross-validation.} We compare our methods with other unsupervised anomaly segmentation baselines, tumor synthesis strategies, and fully-supervised methods. Our method significantly outperforms all other state-of-the-art unsupervised baseline methods and even surpasses the fully-supervised method with detailed \textit{pixel-wise annotation}.
    }
    \label{tab:synt_real_result}
\end{table*}

\smallskip\noindent\textbf{\textit{Evaluation metrics.}}
Tumor segmentation performance was evaluated by Dice similarity coefficient (DSC) and Normalized Surface Dice (NSD) with 2mm tolerance; tumor detection performance was evaluated by Sensitivity and Specificity.
For all the metrics above, 95\% CIs were calculated and the \textit{p}-value cutoff of less than 0.05 was used for defining statistical significance.

\smallskip\noindent\textbf{\textit{Implementation.}} Our codes are implemented based on the MONAI\footnote{\href{https://monai.io/}{https://monai.io/}} framework for both U-Net and Swin UNETR. Input images are clipped with the window range of [-21,189] and then normalized to have zero mean and unit standard deviation. Random patches of $96\times96\times96$ were cropped from 3D image volumes during training. 
All models are trained for 4,000 epochs, and the base learning rate is 0.0002. The batch size is two per GPU. 
We adopt the linear warmup strategy and the cosine annealing learning rate schedule. 
For inference, we use the sliding window strategy by setting the overlapping area ratio to 0.75.

\begin{figure*}[t]
\centerline{\includegraphics[width=1.0\linewidth]{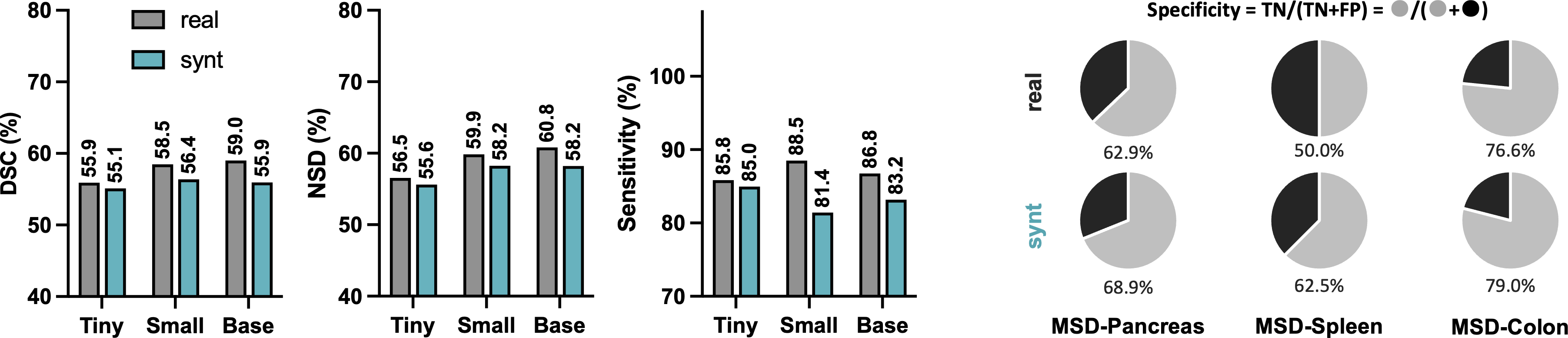}}
    \caption{
    \textbf{Generalization to different models and data.}
     Training U-Net on synthetic liver tumors outperforms as well as training it on real tumors with per-voxel annotation (see~\tableautorefname~\ref{tab:synt_real_result}). We further examine this observation using Swin UNETR~\cite{tang2022self}, including its variance of Tiny (Param = 4.0M), Small (Param = 15.7M), and Base (Param = 62.1M). The DSC, NSD, and Sensitivity scores are evaluated on the LiTS datasets. The detailed results of 5-fold cross-validation are reported in Appendix~\tableautorefname~\ref{tab:fold_evaluation}. Moreover, the model trained on synthetic tumors can also be generalized to CT scans with the \textit{healthy} liver across datasets (\eg MSD-Pancreas, MSD-Spleen, and MSD-Colon~\cite{antonelli2021medical}), generating fewer false positives and yielding  a higher Specificity compared with the model trained on real tumors. 
    }
\label{fig:tiny_small_base}
\end{figure*}

\begin{figure}[t]
\centerline{\includegraphics[width=1.0\linewidth]{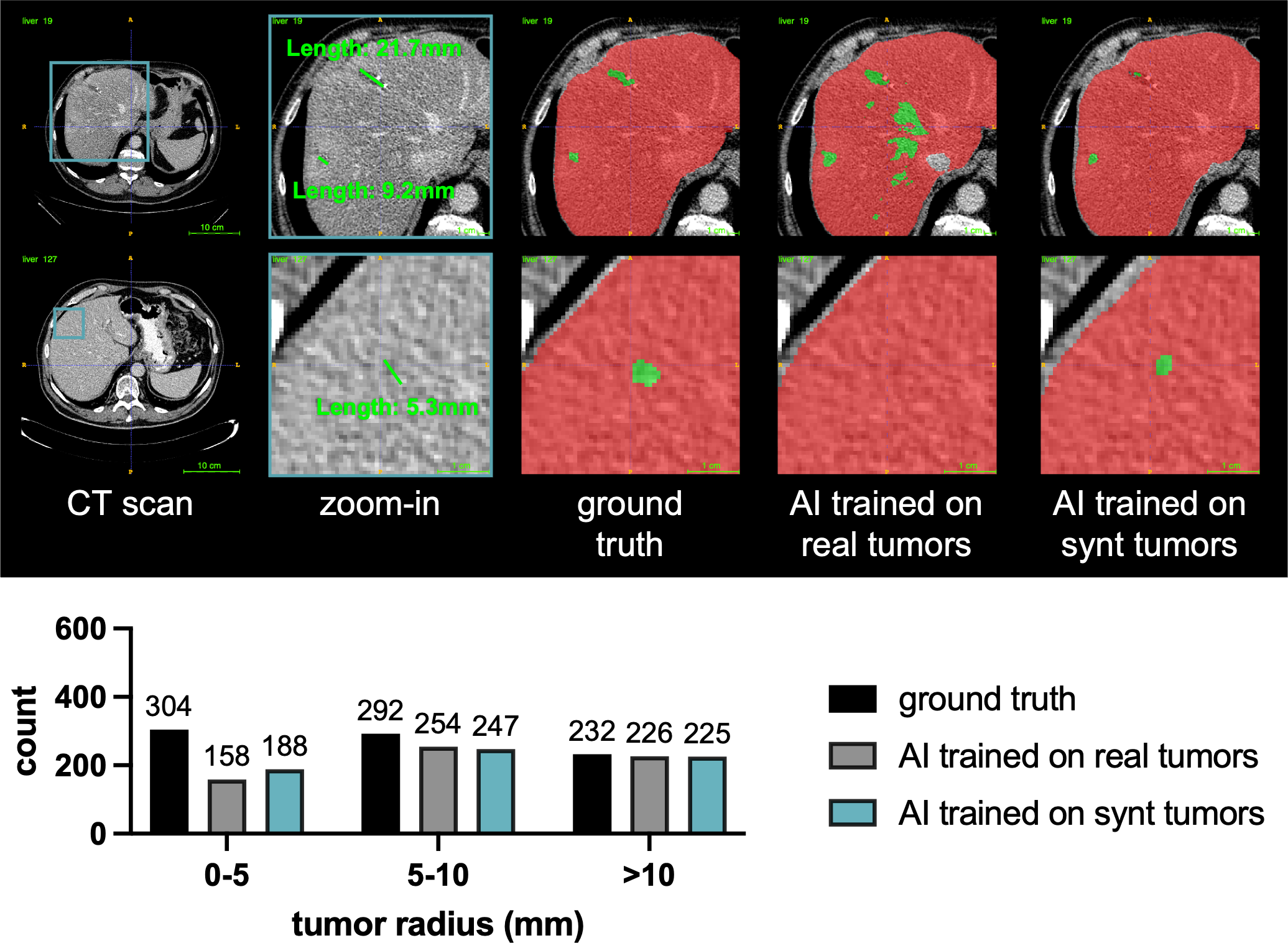}}
    \caption{
        \textbf{Small tumor detection.}
        The upper panel presents two examples of small tumors and the segmentation results. For both models trained on real and synthetic tumors, the false negatives are mostly smaller than 10mm. 
        The lower panel presents the tumor detection rate. The model trained on synthetic tumors could detect tumors as small as 2mm.
    }
\label{fig:performance_tumor_size}
\end{figure}

\begin{figure*}[!h]
    \centering
    \includegraphics[width=1.0\linewidth]{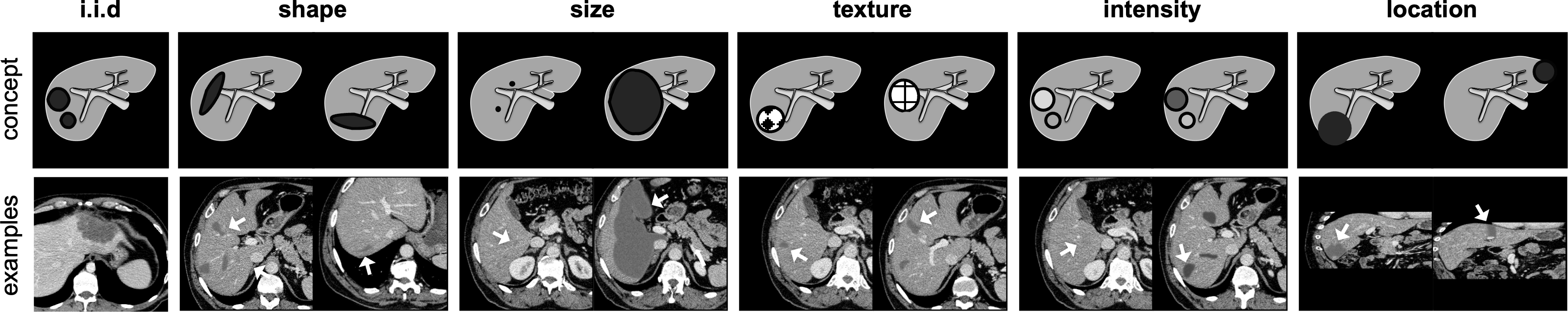}
    \vfill
    \vspace{0.05em}
    \scriptsize
    \vspace{1em}
    \begin{tabular}{p{0.12\linewidth}|p{0.032\linewidth}p{0.032\linewidth}p{0.04\linewidth}|p{0.032\linewidth}p{0.032\linewidth}p{0.032\linewidth}|p{0.032\linewidth}p{0.032\linewidth}p{0.032\linewidth}|p{0.032\linewidth}p{0.032\linewidth}p{0.032\linewidth}|p{0.032\linewidth}p{0.032\linewidth}p{0.032\linewidth}}
        \hline
         & \multicolumn{3}{c|}{shape} & \multicolumn{3}{c|}{size} & \multicolumn{3}{c|}{texture} & \multicolumn{3}{c|}{intensity} & \multicolumn{3}{c}{location} 
         \\
         & $\mu$$\pm$$\sigma$ & $\mu$$\pm$2$\sigma$ & $\mu$$\pm$3$\sigma$ & $\mu$$\pm$$\sigma$ & $\mu$$\pm$2$\sigma$ & $\mu$$\pm$3$\sigma$ & $\mu$$\pm$$\sigma$ & $\mu$$\pm$2$\sigma$ & $\mu$$\pm$3$\sigma$ & $\mu$$\pm$$\sigma$ & $\mu$$\pm$2$\sigma$ & $\mu$$\pm$3$\sigma$ & $\mu$$\pm$$\sigma$ & $\mu$$\pm$2$\sigma$ & $\mu$$\pm$3$\sigma$ 
        \\
        \shline
        UNet++~\cite{zhou2019unet++} & 81.84 & 85.78 & 84.35 & 68.45 & 63.01 & 9.27 & 75.92 & 85.75 & 82.54 & 90.16 & 75.58 & 26.99 & 84.12 & 83.89 & 81.49
        \\
        nnU-Net~\cite{isensee2021nnu} & 82.18 & 83.85 & 85.44 & 80.23 & 59.55 & 5.39 & 84.91 & 88.47 & 84.18 & 91.60 & 83.61 & 30.53 & 84.84 & 85.42 & 84.06
        \\
        Swin UNETR~\cite{tang2022self} & 81.79 & 81.82 & 82.37 & 82.62 & 65.95 & 26.08 & 85.43 & 86.12 & 82.31 & 88.95 & 79.36 & 12.87 & 84.05 & 82.71 & 80.23
        \\
        \hline
    \end{tabular}
\caption{
    \textbf{Controllable benchmark for robust evaluation.} UNet++~\cite{zhou2018unet++}, nnU-Net~\cite{isensee2021nnu}, and Swin UNETR~\cite{tang2022self} are very competitive segmentation models in the medical domain. However, their limitations of tumor segmentation are not fully revealed due to the lack of sufficient testing images (only 70 CT scans are available for testing in MSD Task03~\cite{antonelli2022medical}). On the contrary, synthetic tumors enable us to perform an extensive evaluation of these models in segmenting liver tumors that vary from different conditions, \ie shape, size, texture, intensity, and location. We downloaded the checkpoints of these models trained on LiTS, and evaluated them on synthetic tumors. Synthetic tumors were generated using different parameters, where $\mu$ and $\sigma$ denote mean and standard deviation, respectively. Our findings indicate that these public models exhibit robustness to variations in tumor shape, location, and texture, but they are sensitive to tumor size and intensity. Specifically, these models are prone to errors when encountering tumors that are smaller or larger than those in the training set, or when faced with differing Hounsfield Unit (HU) values, which may be attributable to contrast enhancement.
}
\label{fig:ood_evaluation}
\end{figure*}

\section{Results \& Discussion}
\label{sec:result}

\noindent\textit{Using 116 CT scans from Pancreas-CT, CHAOS, and BTCV with our label-free tumor generator, we outperformed all those methods on the LiTS benchmark, wherein previous methods used 101 CT scans and annotations from LiTS.}

\subsection{Clinical Validation using Visual Turing Test}
\label{sec:visual_turing_test}

We conduct the Visual Turing Test~\cite{geman2015visual} on 50 CT scans, where 20 scans are with real tumors from LiTS, and the remaining 30 scans are healthy livers from WORD~\cite{luo2022word} with synthetic tumors. Two professionals with different experience levels take part in this test. They can inspect each sample in 3D view, which means continuous variation of slice sequence can be observed by scrolling the mouse. This is an important setting for the test because some important tumor characteristics are not obvious in a 2D view (\eg, vessel collision). In the test, professionals can label each sample as \textit{real}, \textit{synthetic} or \textit{unsure}. When calculating the performance metrics, only the samples with definite results are counted.

The testing results are shown in \tableautorefname~\ref{tab:turing_test_result}. For junior professionals with 6-year experience, definite judgments of 49 out of 50 samples are given. All of the accuracy, sensitivity, and specificity are below 30\%, which means the generated samples succeed in confusing the junior professional. In particular, the sensitivity of 27.6\% means that the rest 72.4\% synthetic samples are mistakenly regarded as real samples. The result verifies that our synthesis method can generate realistic tumors. According to the results given by the senior professional with 15-year experience, 36.8\% synthetic samples seemed to be real, indicating that nearly half of the generated samples can tease senior professionals. Noteworthy, the senior professional only gives 19 judgments among all 30 synthetic samples. Adding up misjudged samples and uncertain samples, a total of 18 out of 30 generated samples have confused him/her.

\subsection{Comparison with State-of-the-art Methods}\label{sec:outperform_baseline}

We compare our label-free tumor synthesis strategy with several prominent unsupervised tumor segmentation methods designed for both natural and medical images, such as PatchCore~\cite{roth2022towards}, f-AnoGAN~\cite{schlegl2019f}, VAE~\cite{baur2021autoencoders}, and the method proposed by Yao~\etal~\cite{yao2021label}. To enhance the performance of these baseline methods, we focus solely on the liver region for training and testing to minimize noise caused by extraneous information. \tableautorefname~\ref{tab:synt_real_result} shows that all the previous unsupervised methods exhibit suboptimal performance in segmenting real liver tumors. In contrast, our label-free tumor synthesis---a novel approach to unsupervised tumor segmentation---significantly outperforms all these methods, achieving a DSC of 59.77\% and an NSD of 61.29\%. On the other hand, the model trained on real tumors using fully supervised learning achieves a DSC of 57.51\% and an NSD of 58.04\%. These results highlight the potential of a paradigm shift from \textit{label-intensive} to \textit{label-free} tumor segmentation.

\subsection{Generalization to Different Models and Data}
\label{sec:generalization_sim2real}

We verify the generalizability of synthetic tumors using Swin UNETR\footnote{Swin UNETR is a hybrid segmentation architecture, which integrates the benefits of both U-Net~\cite{ronneberger2015u} and Transformer~\cite{dosovitskiy2020image,liu2021swin}. We select Swin UNETR because it is very competitive and has ranked first in numerous public benchmarks~\cite{tang2022self}, including liver tumor segmentation (MSD-Liver).}~\cite{hatamizadeh2022swin}, including its Tiny, Small, and Base variants. \figureautorefname~\ref{fig:tiny_small_base} shows that the model trained on real tumors performs slightly better than that on synthetic tumors, but there is no statistical difference between the two results as the \textit{p}-value is greater than 0.05.
In addition to evaluating the models on the LiTS dataset, we assess their domain generalization ability using data from other datasets\footnote{We first selected the tumor-free scans from these datasets and then had radiologists review each one of the scans to dismiss the diseased liver.} (\ie, MSD-Pancreas, MSD-Spleen, MSD-Colon). As shown in the right panel of \figureautorefname~\ref{fig:tiny_small_base}, our model trained with healthy data collected from 3 different datasets shows better robustness than the model trained on real data only from LiTS, while achieving much higher Specificity on the three external datasets. It is noteworthy that higher Specificity (fewer false positives) is crucial in clinical applications as it reduces the number of patients subjected to invasive diagnostic procedures and their associated costs~\cite{brodersen2013long,xia2022felix,liu2023clip}.

\subsection{Potential in Small Tumor Detection}
\label{sec:small_tumor_detection}
Early detection of small tumors is essential for prompt cancer diagnosis, but such cases are scarce in real datasets because most patients remain asymptomatic during the early stages. AI models trained on these datasets exhibit reduced detection sensitivity for small tumors (radius $<$ 5mm) compared with larger tumors (radius $>$ 5mm), displaying sensitivities of 52.0\% and 91.6\%, respectively. Thus, an advanced tumor generator could create synthetic data containing various tumor sizes for training and testing models, addressing the size imbalance issue found in real data.
The lower panel of \figureautorefname~\ref{fig:performance_tumor_size} presents quantitative tumor detection performance stratified by tumor size, and the upper panel presents two cases with small tumors for qualitative comparison. Evidently, AI models (trained solely on synthetic data) outperform those trained on real tumors in detecting and segmenting small tumors in the liver. The results indicate that the generation of copious amounts of synthetic small tumors can improve the efficacy of models in detecting real small tumors, thereby playing a crucial role in the early detection of cancer.

\subsection{Controllable Robustness Benchmark}
\label{sec:robustness_benchmark}

Standard evaluation in medical imaging is limited to determining the effectiveness of AI in detecting tumors. This is because the number of annotated tumors in the existing test datasets is not big enough to be representative of the tumors that occur in real organs and, in particular, contains only a limited number of very small tumors. We show that synthetic tumors can serve as an accessible and comprehensive source for rigorously evaluating AI's performance in detecting tumors at a variety of different sizes and locations with the organs. 
To be specific, our tumor generator can synthesize liver tumors varying in five dimensions, \ie location, size, shape, intensity, and texture, by tuning hyperparameters in the tumor generation pipeline. Taking five different options in each dimension, our tumor generator could create 25 (5$\times$5) variants for each single CT scan. 
Generating a large number of synthetic tumors during testing enables us to find failure scenarios of current AI models. After locating the worst cases of AI, we can synthesize and include worst-case tumors in the training set and fine-tune AI algorithms.
\figureautorefname~\ref{fig:ood_evaluation} illustrates the out-of-distribution (o.o.d.) benchmark created by synthetic tumors, wherein we evaluate several state-of-the-art models (trained on public datasets). These models show good robustness in the shape, location, and texture dimensions but are sensitive to tumors of extreme sizes and intensities.

\begin{table}[t]
\centering
\footnotesize
\setlength{\tabcolsep}{1.6mm}
\renewcommand{\arraystretch}{1.0}
\begin{tabular}{ccc|cc}
\hline
tiny size             & \begin{tabular}[c]{@{}c@{}}elastic\\deformation\end{tabular}
& edge blurring              & \begin{tabular}[c]{@{}c@{}} all tumors \\ DSC (\%) \end{tabular}   & \begin{tabular}[c]{@{}c@{}} small tumors \\ det Sen. (\%) \end{tabular} \\ \shline

\cmark &          &                     & 43.9 & 26.4               \\
\cmark &        & \cmark                & 47.1 & 51.3               \\
\cmark & \cmark &                             & 50.3 & 54.1                \\
 & \cmark         & \cmark                    & 52.6 & 33.4                \\
\cmark & \cmark         & \cmark                    & 55.1 & 61.8       \\   \hline     
\end{tabular}
\caption{\textbf{Ablation study on shape generation.} The quality of synthesized tumors influences model performance to a certain degree, emphasizing the importance of each component in our proposed method (\S\ref{sec:method}; \figureautorefname~\ref{fig:method}). The quality assessment of generated tumors is in Appendix~\figureautorefname~\ref{fig:model_training_type}. Moreover, generating tiny synthetic tumors positively impacts the sensitivity of small tumors.}
\label{tab:ablation_study}
\end{table}

\subsection{Ablation Study on Shape Generation}\label{sec:ablation_study}

To show the importance of  each step in tumor generation, we design ablation studies focusing on shape generation and synthesizing small tumors. We evaluate the models trained with different incomplete settings of synthetic strategies on two aspects: all tumor segmentation and small tumor detection. As shown in \Cref{tab:ablation_study}, the performance would be much poorer without synthesizing small tumors, edge blurring or elastic deformation in the shape generation. The reasons are simple: (1) without elastic deformation and edge blurring steps for shape generation (shown in Figure~\ref{fig:method}), the synthetic tumors can be extremely unrealistic (\ie the edge is sharp and shape can only be ellipsoid). Several examples are provided in Appendix Figure~\ref{fig:ablation_settings}. (2) The model doesn't have the generalization ability to small tumors (radius $<$ 5mm) when the training set does not have them.

\section{Conclusion}

In this paper, we have developed an effective strategy to synthesize liver tumors. With \textit{zero} manual annotation, we verify that the AI model trained on synthetic tumors can perform similarly to the ones trained on real tumors in the LiTS dataset (which took months to create). This reveals the great potential for the use of synthesis tumors to train AI models on larger-scale healthy CT datasets (which are much easier to obtain than CT scans with liver tumors). Furthermore, synthetic tumors allow us to assess AI's capability of detecting tumors of varying locations, sizes, shapes, intensities, textures, and stages in CT scans. 
In the future, we will consider generative adversarial nets (GANs)~\cite{goodfellow2020generative,han2019synthesizing}, Diffusion Models~\cite{ho2020denoising} and possibly improved with 3D geometry models like NeRF~\cite{mildenhall2021nerf} to generate better tumor texture.



\noindent\textbf{Acknowledgements.} This work was supported by the Lustgarten Foundation for Pancreatic Cancer Research and partially by the Patrick J. McGovern Foundation Award.
We appreciate the effort of the MONAI Team to provide open-source code for the community. We thank Yucheng Tang, Huimiao Chen, Bowen Li, Jessica Han, and Wenxuan Li for their constructive suggestions; thank Camille Torrico and Alexa Delaney for improving the writing of this paper. 
Paper content is covered by patents pending.

\clearpage
{\small
\bibliographystyle{ieee_fullname}
\bibliography{refs,zzhou}
}

\newpage
\clearpage
\appendix


\begin{table*}[!t]
    \centering
    \scriptsize
    \begin{tabular}{p{0.12\linewidth}P{0.15\linewidth}P{0.07\linewidth}P{0.073\linewidth}P{0.073\linewidth}P{0.073\linewidth}P{0.073\linewidth}P{0.073\linewidth}P{0.073\linewidth}}
        \hline
        \textit{\textbf{U-Net}} & labeled/unlabeled CTs & metric & fold 0 & fold 1 & fold 2 & fold 3 & fold 4 & average \\
        \shline
        \multirow{2}{*}{real} & \multirow{2}{*}{101/0} & DSC~(\%) & 55.86 & 52.26 & 67.34 & 53.06 & 59.63 & 57.63
        \\
         & & NSD~(\%) & 56.87 & 49.02 & 68.54 & 55.02 & 61.06 & 58.10
        \\
        \hline
        \multirow{2}{*}{synt} & \multirow{2}{*}{0/116} & DSC~(\%) & 61.83 & 50.38 & 69.63 & 57.75 & 59.46 & 59.81
        \\
         & & NSD~(\%) & 64.50 & 47.74 & 71.48 & 61.96 & 60.22 & 61.28
        \\
        \hline
        \multirow{2}{*}{real \& synt} & \multirow{2}{*}{50/52} & DSC~(\%) & 56.96  & 48.77 & 68.65 & 54.16 & 55.76 &  56.86
        \\
         & & NSD~(\%) & 59.09 & 43.21 & 69.44 & 54.01 & 54.56 & 56.06
        \\
        \hline
    \end{tabular}
    \vfill
    \vspace{1em}
    \begin{tabular}{p{0.12\linewidth}P{0.15\linewidth}P{0.07\linewidth}P{0.073\linewidth}P{0.073\linewidth}P{0.073\linewidth}P{0.073\linewidth}P{0.073\linewidth}P{0.073\linewidth}}
        \textit{\textbf{Swin~UNETR-Tiny}} & labeled/unlabeled CTs & metric & fold 0 & fold 1 & fold 2 & fold 3 & fold 4 & average \\
        \shline
        \multirow{2}{*}{real} & \multirow{2}{*}{101/0} & DSC~(\%) & 52.88 & 49.24 & 67.94 & 53.93 & 55.63 & 55.92 
        \\
         & & NSD~(\%) & 51.34 & 47.08 & 71.22 & 54.56 & 58.53 & 56.55
        \\
        \hline
        \multirow{2}{*}{synt} & \multirow{2}{*}{0/116} & DSC~(\%) & 55.90 & 49.63 & 62.20 & 52.48 & 55.30 & 55.10 
        \\
         & & NSD~(\%) & 59.97 & 46.92 & 63.23 & 54.08 & 53.88 & 55.61
        \\
        \hline
    \end{tabular}
    \vfill
    \vspace{1em}
    \begin{tabular}{p{0.12\linewidth}P{0.15\linewidth}P{0.07\linewidth}P{0.073\linewidth}P{0.073\linewidth}P{0.073\linewidth}P{0.073\linewidth}P{0.073\linewidth}P{0.073\linewidth}}
        \textit{\textbf{Swin~UNETR-Small}} & labeled/unlabeled CTs & metric & fold 0 & fold 1 & fold 2 & fold 3 & fold 4 & average \\
        \shline
        \multirow{2}{*}{real} & \multirow{2}{*}{101/0} & DSC~(\%) & 60.01 & 50.56 & 69.83 & 52.08 & 59.98 & 58.49 
        \\
         & & NSD~(\%) & 64.40 & 48.67 & 71.20 & 55.34 & 59.68 & 59.86
        \\
        \hline
        \multirow{2}{*}{synt} & \multirow{2}{*}{0/116} & DSC~(\%) & 57.16 & 52.16 & 63.63 & 54.79 & 54.13 & 56.37
        \\
         & & NSD~(\%) & 63.61 & 50.04 & 66.89 & 57.66 & 52.98 & 58.24
        \\
        \hline
    \end{tabular}
    \vfill
    \vspace{1em}
    \begin{tabular}{p{0.12\linewidth}P{0.15\linewidth}P{0.07\linewidth}P{0.073\linewidth}P{0.073\linewidth}P{0.073\linewidth}P{0.073\linewidth}P{0.073\linewidth}P{0.073\linewidth}}
        \textit{\textbf{Swin~UNETR-Base}} & labeled/unlabeled CTs & metric & fold 0 & fold 1 & fold 2 & fold 3 & fold 4 & average \\
        \shline
        \multirow{3}{*}{real} & \multirow{3}{*}{101/0} & DSC~(\%)$^{\dagger}$ & \leavevmode\color{lightgray}55.35  & \leavevmode\color{lightgray}50.32  & \leavevmode\color{lightgray}64.41  & \leavevmode\color{lightgray}54.17  & \leavevmode\color{lightgray}55.35  & \leavevmode\color{lightgray}55.92
        \\
         & & DSC~(\%) & 59.19 & 54.04 & 68.32 & 52.58 & 60.97 & 59.02
        \\
         & & NSD~(\%) & 63.56 & 52.46 & 70.06 & 55.19 & 62.85 & 60.82
        \\
        \hline
        \multirow{2}{*}{synt} & \multirow{2}{*}{0/116} & DSC~(\%) & 55.26 & 51.43 & 64.87 & 53.34 & 54.82 & 55.94 
        \\
         & & NSD~(\%) & 62.08 & 49.87 & 67.89 & 57.56 & 53.61 & 58.20
        \\
        \hline
    \end{tabular}
    \begin{tablenotes}
    \scriptsize
        \item $^{\dagger}$The 5-fold cross validation results are provided by Tang~\etal~\cite{tang2022self}.
    \end{tablenotes}
    \caption{
    \textbf{Performance on 5-fold cross-validation.} We compare the model (U-Net, Swin-UNETR-Tiny, Small, Base) trained on synthetic tumors with the model trained on real tumors with 5-fold cross-validation. We use Dice Similarity Coefficient (DSC) and Normalized Surface Distance (NSD) as evaluation metrics to measure tumor segmentation performance. AI models trained solely on synthetic tumors achieve comparable performance to those trained on per-voxel annotation. Furthermore, the U-Net architecture can even exceed the performance of per-voxel annotation. The results indicate that synthetic tumors have the potential to serve as an alternative to real tumors for training AI models. This also signifies a paradigm shift in liver tumor segmentation, transitioning from a label-intensive AI development to a label-free one.}
    \label{tab:fold_evaluation}
\end{table*}

\clearpage
\begin{figure*}[t]
\centerline{\includegraphics[width=1.0\linewidth]{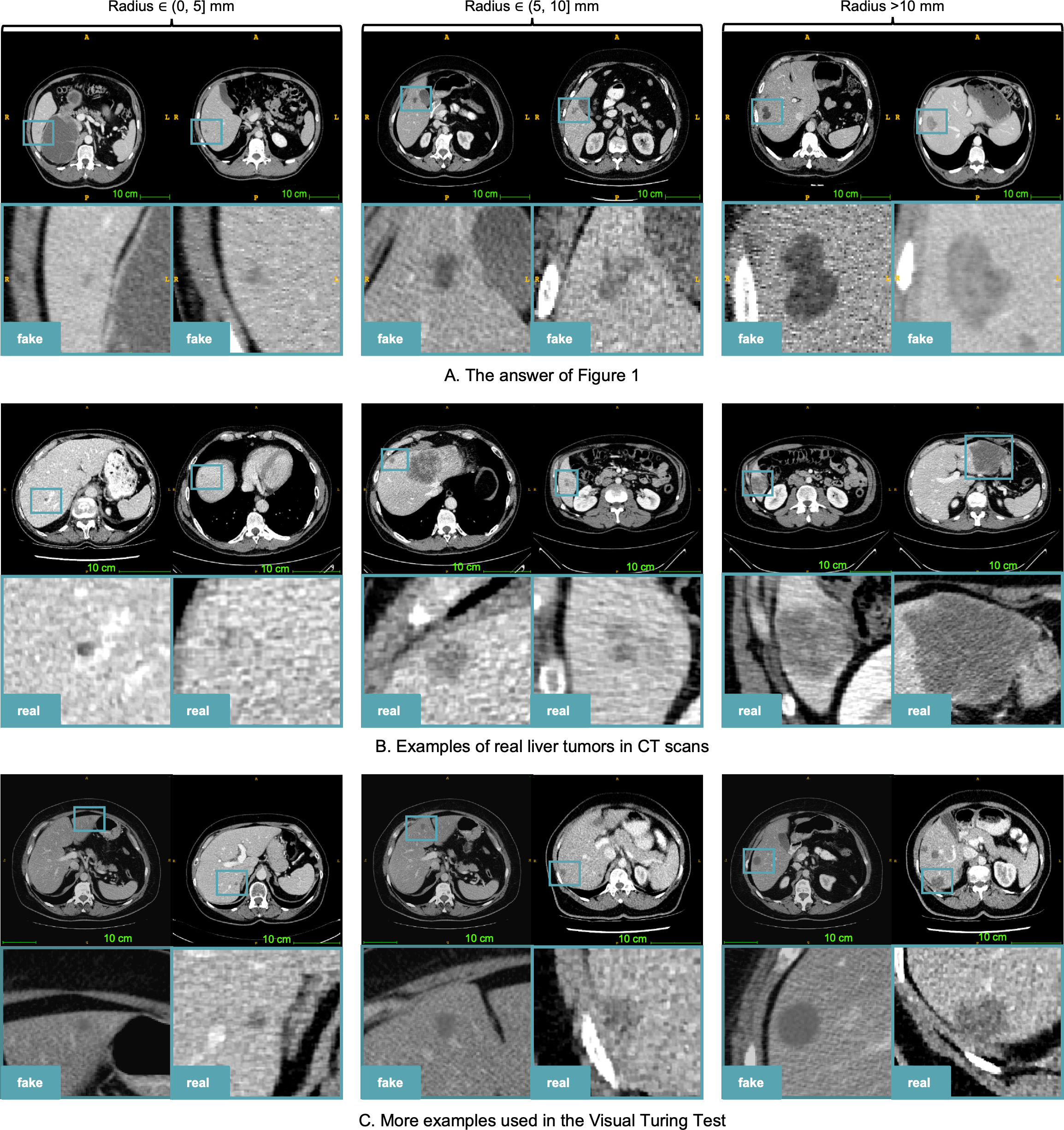}}
    \caption{ \textbf{The answer of \figureautorefname~\ref{fig:teaser}}.
    \textbf{A.} All the six examples in~\figureautorefname~\ref{fig:teaser} are synthetic liver tumors generated by our algorithm.
    \textbf{B.} Examples of real liver tumors stratified by tumor size (small, medium, large).
    \textbf{C.} Examples of the Visual Turing Test for clinical validation. These CT scans are sent to medical professionals (format as \texttt{nii.gz}). The professionals are asked to mark each CT scan as real, synthetic, or unsure.
    Based on results in \S\ref{sec:visual_turing_test} and \tableautorefname~\ref{tab:turing_test_result}, the senior professional achieves an accuracy of 26.5\% with 1 out of 50 CT scans marked unsure, the junior professional achieves an accuracy of 71.0\% with 19 out of 50 marked unsure.
    }
\label{fig:teaser_answer}
\end{figure*}

\clearpage
\begin{figure*}[t]
\centerline{\includegraphics[width=1.0\linewidth]{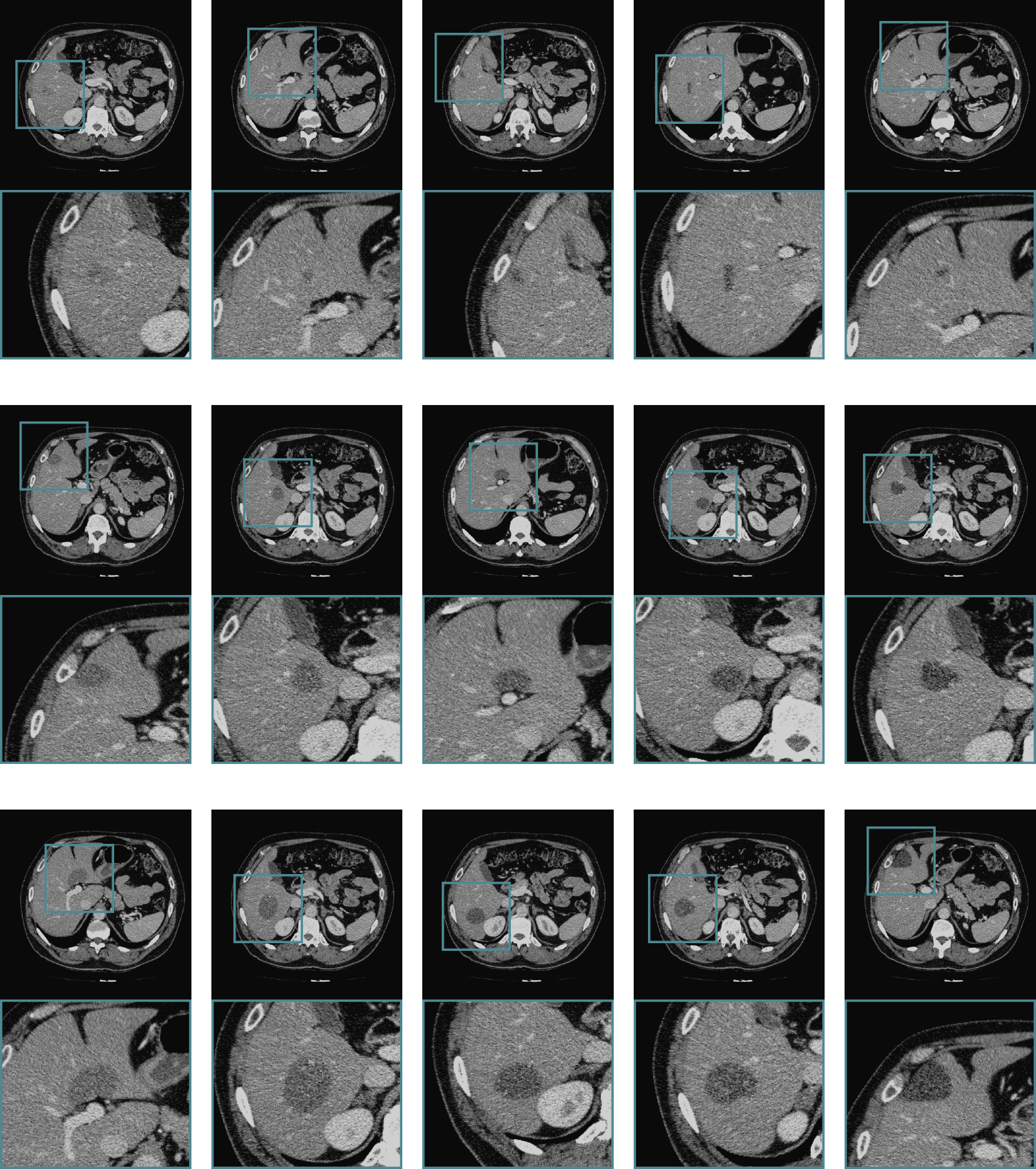}}
    \caption{\textbf{Visualization of tumor generation: examples.}
    We have developed a hand-craft strategy to generate synthetic liver tumors. Our synthetic tumors are realistic in shape and texture, which even medical professionals can confuse with real tumors. On the other hand, the generation pipeline is quite flexible, we can control its shape, size, texture, intensity, and location. This figure shows some examples of synthetic tumors generated by our method. The size of the synthetic tumor exhibits an increase from top to bottom, and its intensity becomes darker from left to right.
    }
\label{fig:extensive_samples}
\end{figure*}

\clearpage
\begin{figure*}[t]
\centerline{\includegraphics[width=0.9\linewidth]{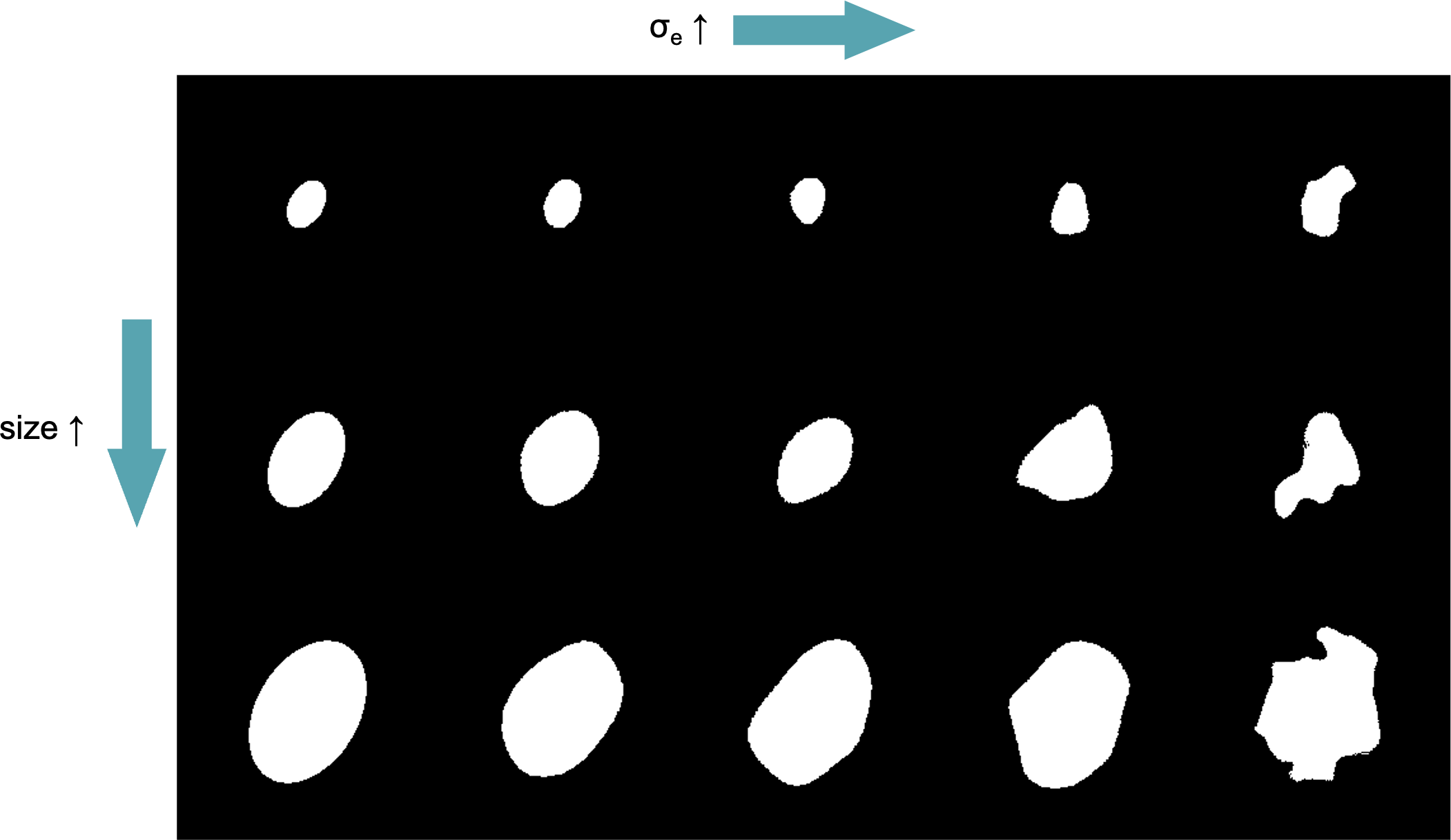}}
    \caption{\textbf{Visualization of tumor generation: shape.} We show the effect of parameters in ``Mask Shape Generation'' (Figure~\ref{fig:method}). The mask shape is controlled by the size $r$ and deformation $\sigma_e$. With the increase of $r$ and $\sigma_e$, the tumor mask shape becomes larger and more irregular. By choosing appropriate numbers, we are able to simulate real tumor shapes.}
\label{fig:shape_parameters}
\end{figure*}

\begin{figure*}[t]
\centerline{\includegraphics[width=0.95\linewidth]{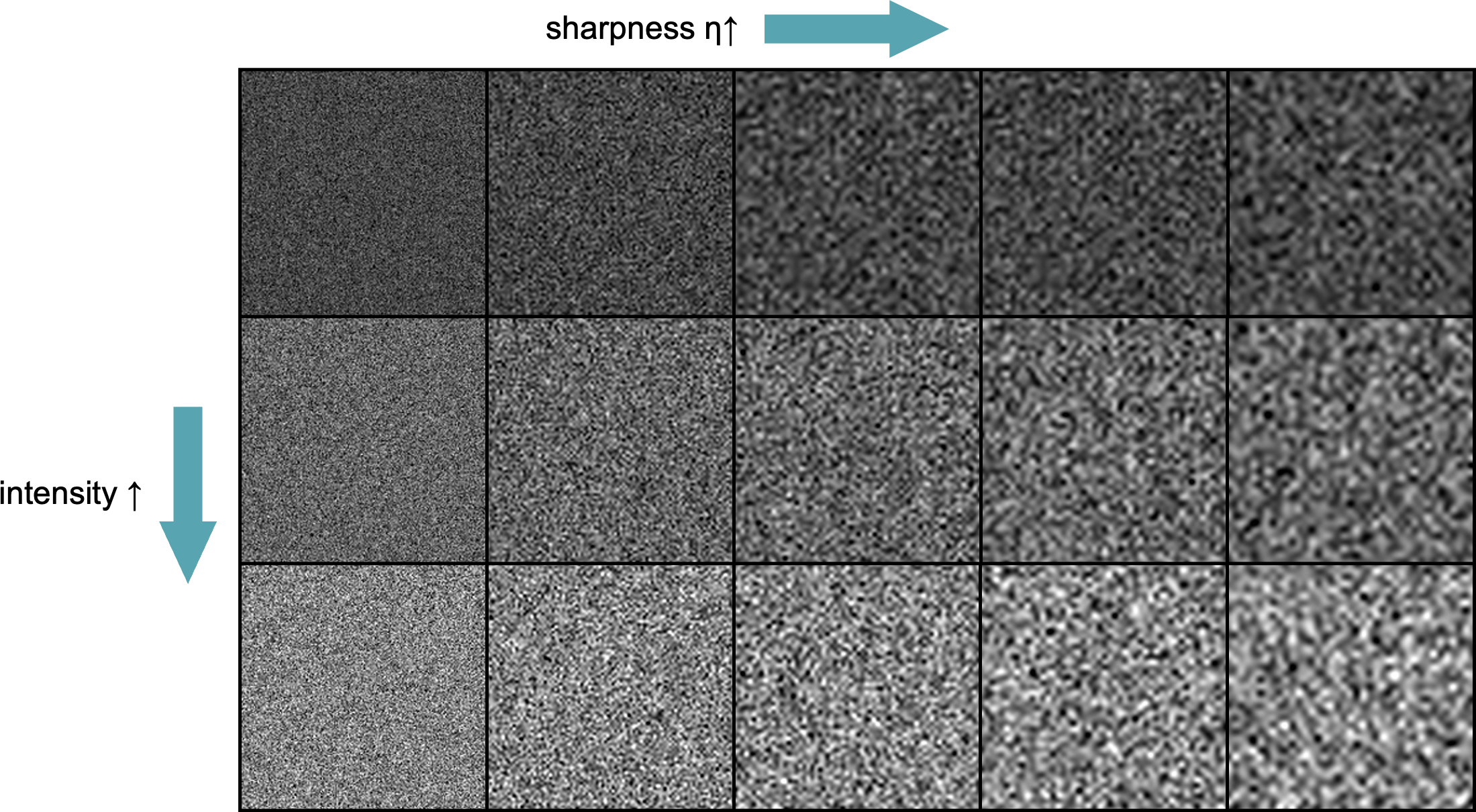}}
    \caption{\textbf{Visualization of tumor generation: texture.} We show the effect of parameters in ``Texture Generation'' (Figure~\ref{fig:method}). The texture of our synthetic tumor is mainly controlled by the intensity $\mu_t$ and sharpness $\eta$. $\mu_t$ represents our synthetic tumor's mean HU value, and $\eta$ determines how rough the generated texture feels. The hyper-parameters we use to simulate real texture can be found in \tableautorefname~\ref{tab:hyper-parameter}.
    }
\label{fig:texture_parameters}
\end{figure*}

\clearpage

\begin{figure*}[t]
\centerline{\includegraphics[width=0.95\linewidth]{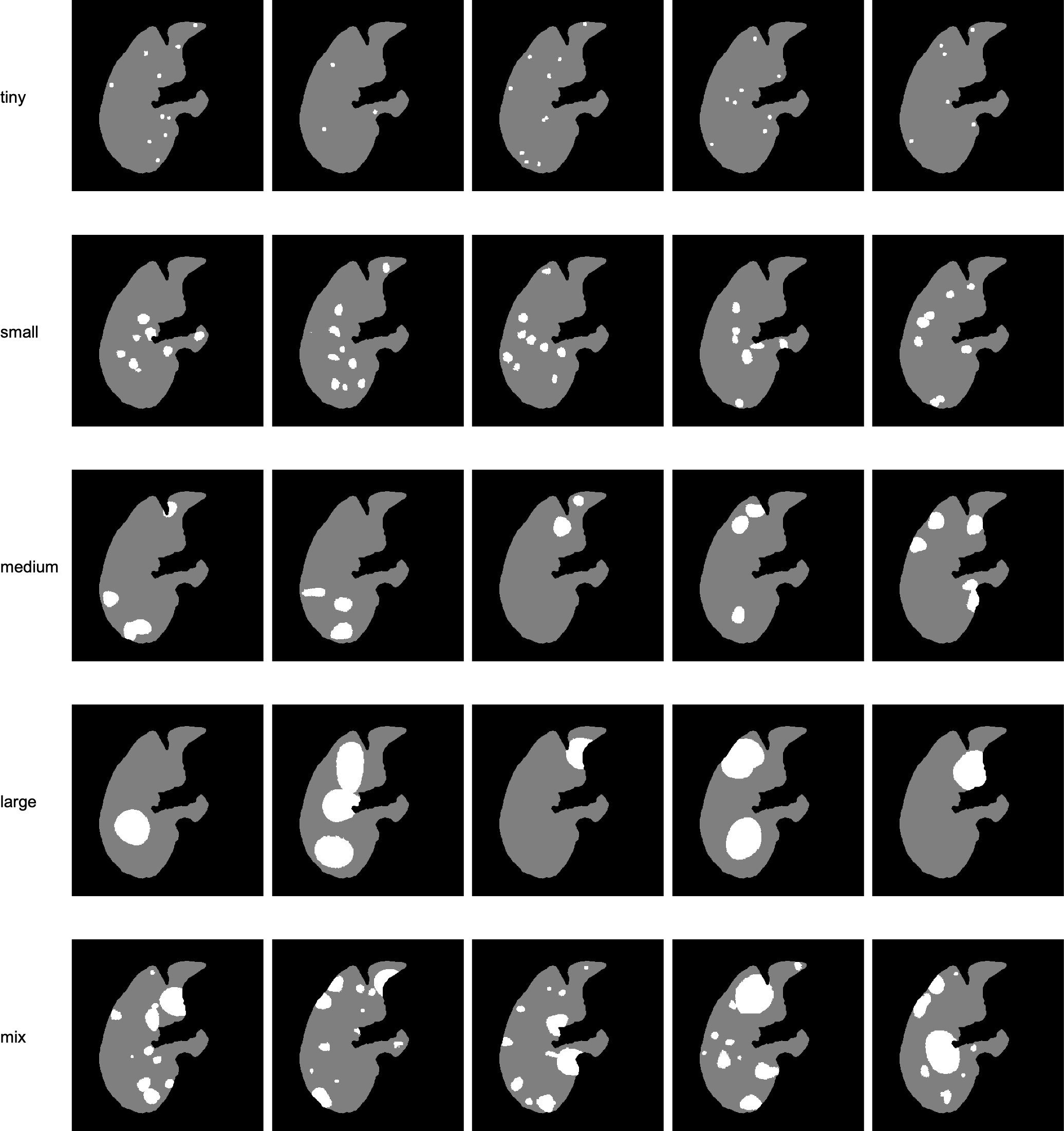}}
    \caption{\textbf{Visualization of tumors for model training.} During training time, we are able to generate liver tumors on the fly, theoretically creating infinite image-label pairs. We show some visualization examples of ``tiny'', ``small'', ``medium'', ``large'', and ``mix'' tumors. The parameters of these tumors are shown in \tableautorefname~\ref{tab:tumor_size_for_training}.
    }
\label{fig:model_training_type}
\end{figure*}

\begin{table*}[t]
\centering
\small
\setlength{\tabcolsep}{5mm}
\renewcommand{\arraystretch}{1.0}
\begin{tabular}{c|ccccc}
\hline
parameter   & tiny & small & medium & large & mix \\ \shline
size r      & 4                        & 8     & 16     & 32    & /   \\
deformation $\sigma_e$ & $U\left[0.5, 1\right]$  & $U\left[1, 2\right]$  &  $U\left[3, 6\right]$  & $U\left[5, 10\right]$      & /   \\
number $N$      & $F\left[3, 10\right]$   & $F\left[3, 10\right]$ &  $F\left[2, 5\right]$  & $F\left[1, 3\right]$      & /  \\
\hline
\end{tabular}
\caption{\textbf{Tumor parameters for model training.} Let $U\left[a, b\right]$ denotes a uniform distribution, $F\left[a, b\right]$ denotes a discrete uniform distribution, $N$ denotes synthetic numbers. To train an AI model, we design 5 different types of tumor sizes, tiny, small, medium, large, and mix combine all. The sample probability during training is [0.2, 0.2, 0.2, 0.2, 0.2], respectively. }
\label{tab:tumor_size_for_training}
\end{table*}

\clearpage
\begin{figure*}[t]
\centerline{\includegraphics[width=0.88\linewidth]{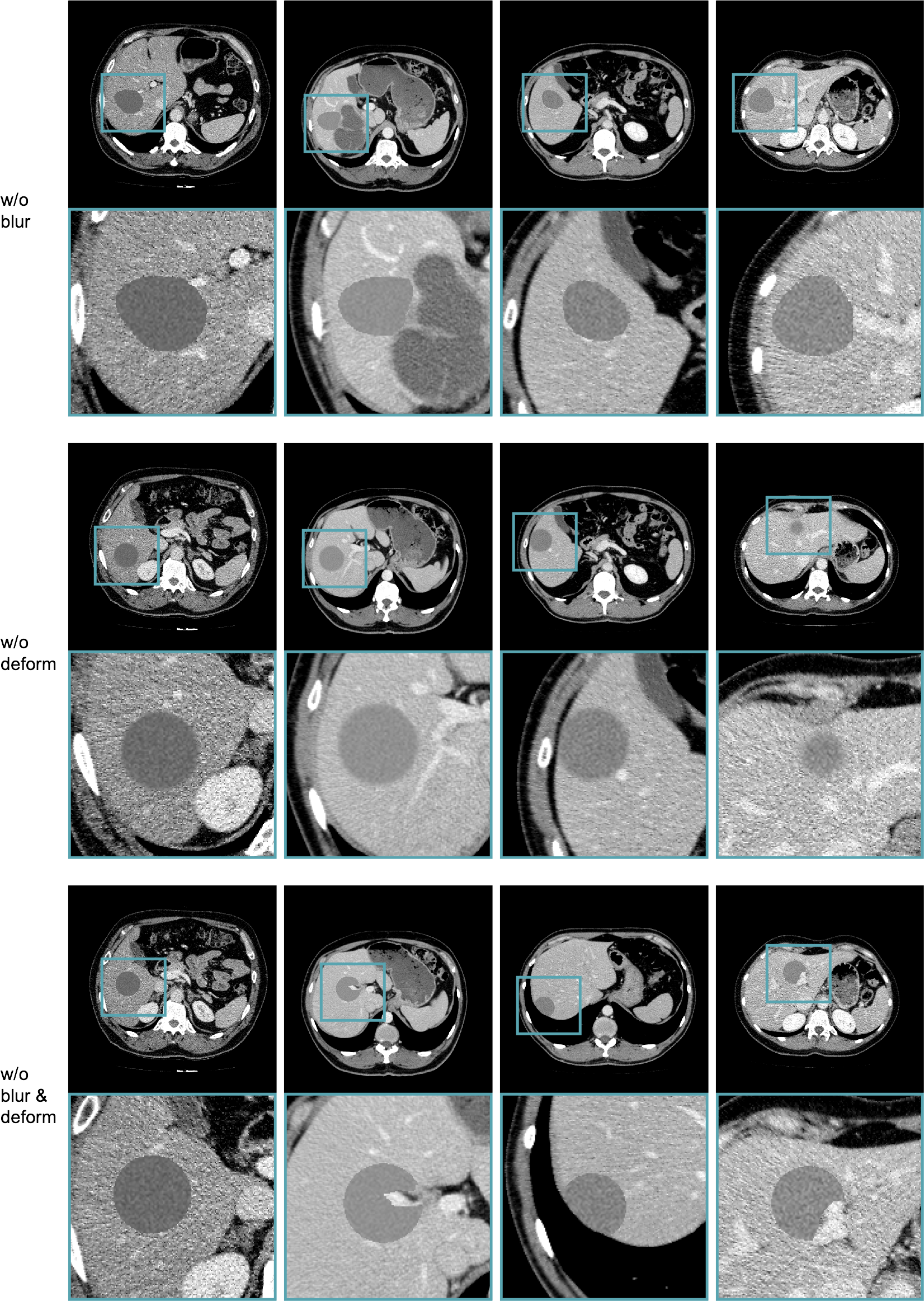}}
    \caption{
    \textbf{Visualization of shape ablation.} To show the importance of synthetic shape, we design ablation studies on ``Mask Shape Generation'' (Figure~\ref{fig:method}). Without edge blurring and elastic deformation, the edge is sharp and the shape can only be ellipsoid. Therefore, synthetic tumors can be extremely unrealistic.
    }
\label{fig:ablation_settings}
\end{figure*}

\end{document}